\documentstyle[epsfig]{mn}
\begin{document}

\title
[Differential magnification of high-$z$ ULIRGS]
{The differential magnification of high-redshift ultraluminous infrared galaxies}
\author 
[A.\,W. Blain] 
{A.\,W. Blain \\
Cavendish Laboratory, Madingley Road, Cambridge, CB3 0HE.\\
Observatoire Midi-Pyr\'en\'ees, 14 Avenue E. Belin, 31400 Toulouse, France.
} 
\maketitle

\begin{abstract}
A class of extremely luminous high-redshift galaxies has recently been 
detected in unbiased submillimetre-wave surveys using the Submillimetre 
Common-User Bolometer Array (SCUBA) camera at the James Clerk Maxwell 
Telescope. Most of the luminosity of these galaxies is emitted from warm 
interstellar dust grains, and they could be the high-redshift counterparts of 
the low-redshift ultraluminous infrared galaxies (ULIRGs). Only one -- 
SMM\,J02399$-$0136 -- has yet been studied in detail. Three other very luminous 
high-redshift dusty galaxies with well determined spectral energy distributions 
in the mid-infrared waveband are known -- IRAS\,F10214+4724, 
H1413+117 and APM\,08279+5255. These were detected serendipitously 
rather than in unbiased surveys, and are all gravitationally lensed by a 
foreground galaxy. Two -- H1413+117 and APM\,08279+5255 -- appear to 
emit a significantly greater fraction of their luminosity in the 
mid-infrared waveband, compared with both low-redshift ULIRGs and 
high-redshift submillimetre-selected galaxies. This can be explained by a 
systematically greater lensing magnification of hotter regions of the source 
compared with cooler regions: differential magnification. This effect 
can confuse the interpretation of the properties of distant ultraluminous 
galaxies that are lensed by intervening galaxies, but offers a possible way to 
investigate the temperature distribution of dust in their nuclei on scales of 
tens of pc.
\end{abstract} 


\section{Introduction} 

At wavelengths between about 10 and 1000\,$\mu$m the spectral energy 
distribution (SED) of a galaxy is 
dominated by the thermal emission from interstellar dust, which typically peaks 
at a wavelength of about 100\,$\mu$m (Sanders \& Mirabel 1996). The dust is 
heated by absorbing the blue and ultraviolet light from young stars and active 
galactic nuclei (AGN). 100-$\mu$m radiation from galaxies at redshifts less 
than about unity has been detected directly by the {\it IRAS} and {\it ISO} 
space-borne telescopes. The redshifted dust emission from more distant 
galaxies can be detected very efficiently at longer submillimetre wavelengths 
(Blain \& Longair 1993a), as demonstrated by the discovery of 
SMM\,J02399$-$0136 at redshift $z=2.8$ 
(Ivison et al.\ 1998; Frayer et al.\ 1998) in a 850-$\mu$m 
survey (Smail, Ivison \& Blain 1997). Distant galaxies were also 
detected by {\it IRAS}, but only those with flux densities enhanced by 
gravitational lensing galaxies. 
Three such galaxies are known -- IRAS\,F10214+4724 
at $z=2.3$ (Rowan-Robinson et al.\ 1991), H1413+117 at $z=2.6$ 
(Barvainis et al.\ 1995) 
and APM\,08279+5255 at $z=3.9$ (Irwin et al.\ 1998; Lewis et al.\ 1998b; 
Downes et al. 1999). The SEDs of these high-redshift galaxies, and two
well studied low-redshift ultraluminous infrared galaxies (ULIRGs) -- Arp 220 
and Markarian 231 (Klaas et al.\ 1997; Rigopoulou, Lawrence \& 
Rowan-Robinson 1996; Soifer et al.\ 1999) -- are shown in Fig.\,1. 

\begin{figure*}
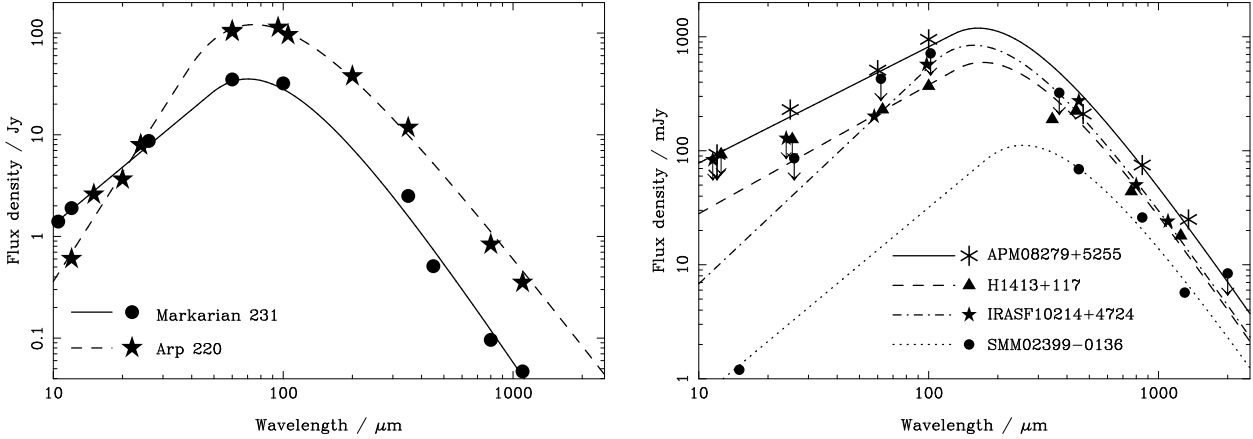

\begin{minipage}{170mm} 
\begin{center}
\epsfig{file=lowz.ps, width=5.75cm, angle=-90}
\hskip 5mm
\epsfig{file=highz.ps, width=5.75cm, angle=-90}
\end{center} 
\caption{The SEDs of two low-redshift (left panel) 
and four high-redshift (right panel) ULIRGs. Note that the mid-infrared SEDs of 
the high-redshift galaxies APM\,08279+5255, H1413+117 and 
IRAS\,F10214+4724 are probably 
modified by differential magnification. The SED models (lines) 
are described by a power-law, 
$f_\nu \propto \nu^a$, at short wavelengths and by a blackbody spectrum at 
temperature $T_{\rm d}$, modified by a dust emissivity 
$\epsilon_\nu \propto \nu^\beta$ with $\beta \simeq 1$ at long wavelengths. 
The values of $T_{\rm d}$ and $a$ in the models are listed in Table\,1.}
\end{minipage} 
\end{figure*}

\section{Spectral energy distributions} 

The difference in the slope of the mid-infrared SED of Arp 220 and Markarian 231
is interpreted as evidence for an active galactic nucleus (AGN) in the core of 
Markarian 231. Powerful 
ionizing radiation from an AGN would heat a very small fraction of the dust 
grains in the galaxy to high temperatures near the nucleus, 
increasing the flux density of the 
galaxy at short wavelengths and thus producing a shallower spectrum. There 
have been various attempts to interpret the mid-/far-infrared SEDs of galaxies 
using models of radiative transfer (Granato, Danese \& Franceschini 1996; 
Green \& Rowan-Robinson 1996). However, given that a power-law spectrum 
convincingly accounts for the limited mid-infrared data available, it is 
reasonable 
to avoid such modeling 
and to assume that 
the mass of emitting 
dust in the source at temperatures between $T$ and $T+{\rm d}T$ is 
$m_{\rm T}(T)\, {\rm d}T$. Dust at 
each temperature can be associated with a $\delta$-function spectrum 
$\delta(\nu-\nu_0)$, in which $\nu_0 \simeq (3+\beta) k T / h$ 
(Blain \& Longair 1993b). $k$ and $h$ are the Boltzmann and Planck constants 
respectively and $\beta$ is the index in the function that describes 
the spectral emissivity of dust $\epsilon_\nu \propto \nu^\beta$; 
$\beta \simeq 1$--1.5. The mid-infrared SED of a source at redshift $z$ is 
\begin{equation} 
f_\nu \propto \int_{T_{\rm min}}^{T_{\rm max}} \, m_{\rm T}(T) \, T^{4+\beta} \,
\delta[\nu(1+z) - \nu_0] \, {\rm d}T.
\end{equation}
If $m_{\rm T} \propto T^\alpha$, then the result is a power-law SED 
$f_\nu \propto \nu^a$, with a
spectral index $a=4+\alpha+\beta$. For 
dust temperatures spanning a range from $T_{\rm min} = 40$\,K to 
$T_{\rm max} = 2000$\,K, this emission covers 
a range of wavelengths from about 90 to 1.8\,$\mu$m. The values of $\alpha$ 
associated with the galaxies above are listed in Table\,1. $\alpha \simeq -6$ 
to $-9$, and so only a very small fraction of the dust in these 
galaxies is heated to high temperatures. 

In the case that a significant fraction of dust heating in a galaxy is caused by 
an AGN, it would be reasonable to assume that the dust temperature would 
depend on the distance from the nucleus of the galaxy $r$ as 
$T_{\rm r}(r) \propto r^\eta$. A negative value of $\eta$ would be expected, 
reflecting a higher dust 
temperature in the more intense radiation field closer to the AGN. 
If the mass of dust enclosed in the spherical shell 
between radii $r$ and $r+{\rm d}r$ is $m_{\rm r}(r)\, {\rm d}r$, where 
$m_{\rm r} \propto r^\gamma$, then the mid-infrared SED, 
\begin{equation} 
f_\nu \propto  \int_{r_{\rm min}}^{r_{\rm max}} m_{\rm r}(r) \, 
T_{\rm r}(r)^{4+\beta} \, \delta[\nu(1+z) - \nu_0] \, {\rm d}r, 
\end{equation}
again evaluates to a power law, with a 
spectral index $a = 3+\beta + (\gamma+1)/\eta$. 
Note that  
for a blackbody in equilibrium with an unobscured point source $\eta = -0.5$. 
In this case $\gamma \simeq 2$ is required in order to represent the SEDs 
of Markarian 231 and SMM\,J02399$-$0136 in Fig.\,1 -- the distribution 
expected if the radial density of dust is uniform. A more reasonable value of 
$\eta$ would be less than $-0.5$, reflecting the effects of obscuration. This 
would imply that $\gamma > 2$, and thus that the density of emitting 
dust increases with increasing radius.

\begin{table*}
\caption{The redshift, restframe dust temperature, and mid-infrared spectral 
index $a$ -- where $f_\nu \propto \nu^a$ -- for the galaxies, the spectral 
energy distributions of which are plotted in Fig.\,1. For comparison, the spectral 
index 
$a \simeq -2.0$ to $-2.5$ for samples of low-redshift {\it IRAS} galaxies, either 
those selected at a wavelength of 25\,$\mu$m (Xu et al.\ 1998) or for the most 
luminous selected at 60\,$\mu$m (Sanders \& Mirabel 1996). The index $\alpha$ 
that is required in the dust mass--temperature function 
$m_{\rm T}(T) \propto T^\alpha$, in order to explain the mid-infrared spectral 
index is also listed for each source. Magnification factors of about 20, 35, 80 
and 400 are required to make the 850-, 450-, 200- and 50-$\mu$m flux 
densities of a $z=2.6$ counterpart of Markarian 231 equal to 
those of H1413+117.
}
{\vskip 0.75mm}
{$$\vbox{
\halign {\hfil #\hfil && \quad \hfil #\hfil \cr
\noalign{\hrule \medskip}
Name & Redshift $z$ & Restframe dust & Mid-infrared spectral & Dust mass \cr  
 & & temperature $T_{\rm d}$ / K & index $a$ & function index $\alpha$ \cr  
\noalign{\smallskip \hrule \smallskip}
Arp 220 & 0.02 & 50 & -3.6 & -8.6 \cr
Markarian 231 & 0.03 & 47 & -1.9 & -6.9 \cr
IRAS\,F10214+4724 & 2.3 & 76 & -1.7 & -6.0 \cr
H1413+117 & 2.6 & 75 & -1.1 & -6.7 \cr
SMM\,J02399$-$0136 & 2.8 & 53 & -1.7 & -6.1 \cr
APM\,08279+5255 & 3.9 & 107 & -1.0 & -6.7 \cr
\noalign{\smallskip \hrule}
\noalign{\smallskip}\cr}}$$}
\end{table*}

Two of the three high-redshift galaxies with {\it IRAS} detections -- 
APM\,08279+5255 and H1413+117 -- have a much flatter mid-infrared SED 
as compared with a typical low-redshift {\it IRAS} galaxy or SMM\,J02399$-$0136: 
see Fig.\,1. SMM\,J02399$-$0136 lacks a detection by {\it IRAS}, but its 
mid-infrared SED is constrained by a 15-$\mu$m {\it ISO} measurement of 
Metcalfe: see Ivison et al. (1998) for further details. Given the limited
mid-infrared data for SMM\,J02399$-$0136, it is certainly possible that its 
mid-infrared spectral index could be more negative than the value of $-1.7$ 
listed in Table\,1. An additional high-redshift source, the brightest detected in a 
850-$\mu$m survey of the Hubble Deep Field (HDF; Hughes et al.\ 1998) with a 
flux density of 7\,mJy, has a reported 15-$\mu$m flux density limit of less than 
23\,$\mu$Jy, indicating that its mid-infrared SED is steeper than that of 
SMM\,J02399$-$0136. 

\begin{figure}
\begin{center}
\vskip -5mm
\epsfig{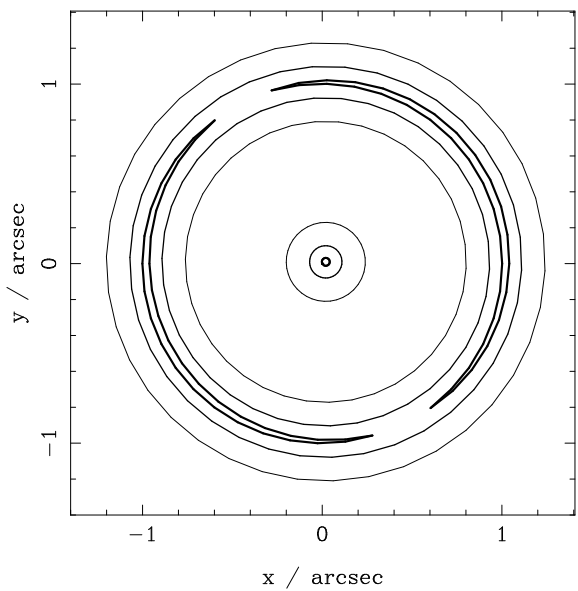}
\end{center}
\caption{An illustration of the effect responsible for differential 
magnification, using a simple model of a lens as a SIS at $z=0.3$ with a 
one-dimensional 
velocity dispersion of 150\,km\,s$^{-1}$. The lens is centred at the origin 
in the figure. The central circular contours mark isophotes that are 
0.5, 2 and 5\,kpc away from the centre of a source at $z=3$, in order of 
decreasing line thickness. The source is centred 0.022\,arcsec from the 
caustic of the lens at the origin. At $z=3$, 1\,arcsec is equivalent to 
11.4\,kpc. The contours at 
radii of about 1\,arcsec in 
the figure correspond to the mapped lensed images of the three source 
isophotes. The magnification for the region of the source enclosed within 
each isophote is 116, 45 and 18 respectively. 
}
\end{figure}

\begin{figure}
\begin{center}
\vskip -5mm
\epsfig{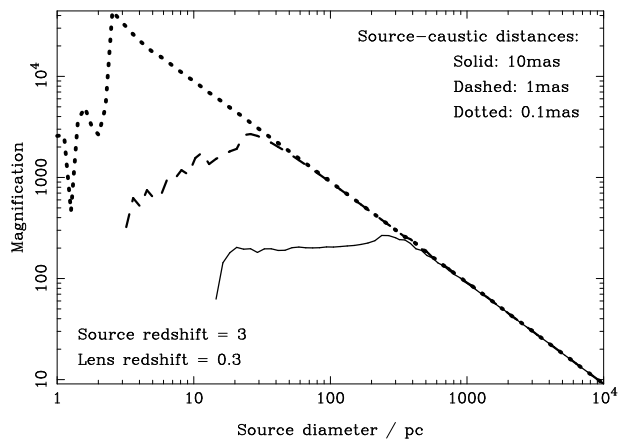}
\end{center}
\caption{The magnification as a function of the source size for the 
SIS lens described in Fig.\,2. The magnification is calculated assuming that 
the source is circular with a uniform surface brightness. 
The solid, dashed and dotted lines show the 
results for a source centred at points with three different offsets from the 
position of the point caustic of the 
SIS lens. The magnifications are large because an SIS lens is assumed. The 
argument in this paper depends on the relative magnification on different
scales and not on the absolute magnification. For a similar plot derived 
for a detailed elliptical lens model 
of IRAS\,F10214+4724, see fig.\,5 of Eisenhardt et al. (1996). The magnification
curves depart from the uniform slope of $-1$ on small scales when the caustic 
point no longer lies within the boundary of the circular source.  
}
\end{figure} 

The slopes of these SEDs are not likely to be 
affected by contaminating sources that are picked up in the differently sized 
observing beams at each wavelength. The beam size of the 0.6-m telescopes 
used to determine the 
mid-infrared SEDs of these galaxies are about 5, 10, 25 and 40\,arcsec at 12, 25, 
60 and 100\,$\mu$m respectively, as compared with about 7 and 14\,arcsec 
for the ground-based submillimetre-wave telescopes used at 450 and 850\,$\mu$m 
respectively. Any contamination from other sources nearby on the sky 
would thus be most  
significant at wavelengths of 60 and 100\,$\mu$m, and lead to a artificial 
steepening of the mid-infrared SED.  

The shallow SEDs~of~APM\,08279+5255 and H1413 +117 could be caused by an 
unusually large fraction of hot dust in these galaxies; however, from optical 
observations of their QSO emission both are known to be gravitationally lensed 
by magnification factors of at least several tens. A systematically greater 
magnification for hotter dust components would increase the flux density at 
shorter wavelengths and so flatten the spectrum (Eisenhardt et al.\ 1996; Lewis 
et al.\ 1998b).  This situation would arise very naturally if the hotter dust 
clouds were smaller and more central, as in the AGN model described in 
equation (2). A similar effect can be produced by the microlensing effect of
individual stars within the lensing galaxy (Lewis et al.\ 1998a). 

For a large magnification to occur, a distant source must lie very close to a 
caustic curve of a gravitational lens. The magnification is formally infinite on 
such a curve, but an upper limit $A_{\rm max}$ is imposed to the magnification 
if the source has a finite size $d$ (Peacock 1982). If the lensing 
galaxy can be modeled as a singular isothermal sphere (SIS), then 
$A_{\rm max} \propto d^{-1}$. Assuming an SIS lens, the image 
geometry and magnifications expected in such a situation are illustrated 
in Figs 2 and 3. Eisenhardt et al. (1996) present more sophisticated lens 
models that account for the geometry of high-resolution images of 
IRAS\,F10214+4724. Ellipticity 
in the lens modifies the magnification--size relation to $A_{\rm max} \propto 
d^{-0.63}$ on scales between 0.001 and 1\,arcsec. In general, 
a similar magnification--size relationship holds regardless of both 
the geometry of the source 
and whether 
one or more objects is responsible for producing the lensing effect 
(Kneib et al.\ 1998). 

The diagnostic feature of such a situation is the 
production of multiple images of comparable brightness. This is clearly the case 
for both APM\,08279+5255 (Lewis et al.\ 1998b) and H1413+117 (Kneib et al.\ 
1998), but less so for IRAS\,F10214+4724, 
which has a more asymmetric arc--counterimage geometry. However, in the lens 
models of both Broadhurst \& L\'ehar (1995) and Eisenhardt et al. (1996), 
IRAS\,F10214+4724 lies very close to the tip of an astroid caustic,  
and so differential lensing would still be expected 
to flatten its mid-infrared SED. These references should be 
consulted for detailed lens models 
of these sources. The lens model if the most recently discovered source 
APM\,08279+5255 is poorly constrained because of the lack of 
high-resolution optical and near-infrared images at present. 

If the smallest, hottest dust clouds that are closest to the AGN 
lie on the caustic that is responsible for the multiple images of the 
AGN (equation 2), then differential magnification of the hot regions compared 
with the cooler regions is likely to modify the SED in the mid-infrared 
waveband. The effect on the SED 
can be calculated by including a magnification factor of 
$A_{\rm max}(r) \propto r^{-\mu}$ in the integrand of equation (2). For a 
SIS lens $\mu = 1$. By evaluating equation (2) in this case, a power-law 
SED with a 
spectral index $a = 3+\beta + (\gamma+1-\mu)/\eta$ is obtained. The spectral 
index is reduced by 
$\Delta a = \mu/\eta$ compared with the value obtained in the 
absence of lensing. 
Note that the 
modification is independent of the value of $\gamma$ and the
form of $m(r)$. Following the same approach, but assuming an exponential 
decrease of $T(r)$, $\Delta a \simeq 1$.

\section{Discussion} 

The mid-infrared SEDs of the high- and low-redshift AGNs SMM\,J02399$-$0136
and Markarian 231 have spectral indices $a \simeq -1.8$. 
Despite being 
lensed by a cluster of galaxies SMM\,J02399$-$0136 does not lie close to a 
caustic, and so is not magnified differentially. Markarian 231 is 
not lensed. 
The strongly lensed 
distant galaxies APM\,08279+5255 and H1413+117 have a spectral index 
$a \simeq -1.1$. This would indicate that $\Delta a = \mu/\eta \simeq -0.7$, and 
thus $\eta \simeq -1.4$ or $-0.9$, depending on whether the lens is a 
SIS, with $\mu=1$,  or matches the model of Eisenhardt et al. (1996), 
with $\mu=0.6$. Both spectral indices are 
considerably steeper than the unscreened blackbody 
value of $-0.5$. Given the very great opacity of interstellar dust to ionizing 
radiation, this is entirely reasonable. To match the observed SEDs, the 
corresponding values of $\gamma$ of 7.3 and 4.0 are required, 
again indicating that most of 
the emitting dust is relatively cool/distant from the core.

IRAS\,F10214+4724 has a steeper mid-infrared SED, and a much smaller optical 
flux density than APM\,08279+5255 and H1413+117 (Lewis et al. 
1998b). Its mid-infrared SED is more similar to those of
Markarian 231 and SMM\,J02399$-$0136. These features could both be
explained if the optical depth of dust extinction into the central regions of 
IRAS\,F10214+4724 were sufficiently large to obscure not only the AGN in the 
optical waveband, but also the hottest dust clouds at longer 
mid-infrared wavelengths. Alternatively, 
the intrinsic unmagnified SED of IRAS\,F10214+4724 could be steeper than 
those of APM\,08279+5255 and H1413+117, just as these two galaxies 
could have intrinsically flat mid-infrared SEDs. 

Does the calculated value of $\eta$ correspond to a plausible size for the 
emitting objects? The temperature of the inner face of the dust cloud exposed 
to the AGN cannot exceed about 2000\,K, or else the dust would sublime. A 
2000-K blackbody would be in equilibrium with a $10^{13}$-L$_\odot$ point 
source at a distance of 0.6\,parsec. If $\eta = -1.4$, then the dust temperature 
falls to 100, 50 and 30\,K at distances of 5, 8 and 14\,parsec respectively. If 
$\eta = -0.9$, then the corresponding distances are 9, 17 and 26\,parsec. These 
values are all in agreement with observations of low-redshift ULIRGs, in which 
the 
emitting region is less than several hundreds of parsecs in extent (Downes \& 
Solomon 1998). Most of the luminosity of the galaxy is emitted by cool dust on 
larger scales. The equivalent radius of a blackbody sphere emitting 
10$^{13}$\,L$_\odot$ at 50\,K is 950\,pc, several times larger than the  
observed sizes of nearby ULIRGs (Downes \& Solomon 1998; Sakamoto et al.\
1999). 

The caustic curves predicted by the lens models that describe IRAS\,F10214+4724 
(Broadhurst \& L\'ehar 1995) and H1413+117 (Kneib et al.\ 1998) are about 0.8 
and 0.2\,arcsec in size in the plane of the source, larger than the scale of a 
200-pc high-redshift source. Hence, the whole emitting region of the source 
should be subject to differential magnification; a difference in magnification
by a factor of about 100 would be obtained between a 200-pc outer radius and a 
2-pc inner radius for an SIS lens. 

If the far-infrared SED of a galaxy is modelled by a single-temperature dust 
spectrum, then differential magnification would be expected to increase the 
temperature for which the best fit was obtained. An increase in temperature by 
a factor of about 20\,per cent would typically be expected for discrete data 
points at the wavelengths given in Fig.\,1. This effect may account for at  
least part of the difference between the very high rest-frame dust temperature 
of about 107\,K inferred for APM\,08279+5255 and the cooler dust temperatures 
inferred for nearby ULIRGs and SMM\,J02399$-$0136. If the redshift of 
a distant differentially magnified ULIRG were to be estimated from an observed 
SED, by fitting to a standard  
template, then the inferred redshift would be underestimated by a similar 
amount, up to 20\,per cent. 
 
The mid-infrared SEDs of distant galaxies that are known to be either unlensed 
or not to be subject to differential magnification are very uncertain.
This is likely to 
remain true for some years, although the {\it Wide-Field Infrared Explorer 
(WIRE)} and {\it Space InfraRed Telescope Facility (SIRTF)} satellites will 
provide some valuable information. Operating at 12 and 25\,$\mu$m, {\it WIRE} 
will specifically probe the mid-infrared SED of distant galaxies. The effect of 
differential magnification discussed 
here could increase the number of lensed AGN detected in these bands 
by a factor of about 10; however, this number is still expected to 
be less than about 1\,per cent of the size of the full {\it WIRE} catalogue. 

AGN-powered ULIRGs are expected to 
have flatter mid-infrared SEDs than starburst-powered ULIRGs, even 
in the absence of differential magnification. Hence, any further flattening due 
to differential magnification should make it easier to distinguish 
AGN from starbursts in the subsample of lensed galaxies in the catalogue. 

The flux density from the inner regions of an 
AGN can experience strong variation on short time-scales. On scales of several 
pc, the response of the hottest, smallest components of the mid-infrared 
dust emission spectrum should be comparable to the light-crossing time -- 
about 1\,yr. Over the lifetime of {\it SIRTF}, any such variations, amplified by 
differential magnification, should be detectable. In order to probe the spatial 
structure and temperature distribution of dust in the inner regions of AGN 
directly, high-resolution mid-infrared imaging, and thus a space-borne 
mid-infrared interferometer, will be required (Mather et al.\ 1998a,b). 
Baselines  
of order 500\,m at 20\,$\mu$m will be required to probe 10-pc scales in 
high-redshift galaxies.  

The conditions in the central regions of AGN can already be probed using a 
variety of methods in other wavebands. Reverberation mapping -- observing
the transient response of line emission to a variable continuum source  -- is 
possible in the optical waveband (Bahcall, Kozlovsky \& Salpeter 1972;
Blandford \& 
McKee 1982; Wandel 1997). Observations of water masers using 
very long baseline radio interferometry (Miyoshi et al. 1995)
have revealed the 
structure of the accretion 
disk in NGC\,4258. 
Modifications to the profiles of 
X-ray fluorescence lines due to strong gravity 
in the innermost regions of accretion disks have also 
been observed  
(Reynolds \& Fabian 1997). 

\section{Conclusions} 

\begin{enumerate} 
\item 
Distant lensed ULIRGs are likely to have their 
mid-infrared SEDs made more shallow by the effects of differential 
magnification. While observations of these sources must be exploited in 
order to investigate the nature of this class of galaxies, careful account must 
be taken of the uncertainties introduced by differential magnification. Recent 
submillimetre-selected samples are expected to be largely immune to this 
problem; their mid-infrared SEDs will be probed by the forthcoming {\it SIRTF} 
mission.  
\item 
The SED of the longest known high-redshift ULIRG IRAS\,F10214+4724 is 
probably affected by differential magnification in the same way as those of 
APM\,08279+5255 and H1413+117. However, it has a steeper mid-infrared 
SED. This could be due to either a steeper intrinsic SED, which is still flattened 
by differential magnification, or to a greater optical depth to dust extinction,
which obscures the most central regions of the galaxy, even in the mid-infrared
waveband.  
\item
The SEDs of a sample of lensed galaxies can be used to probe the conditions 
in the cores of distant dusty galaxies. A larger sample of these objects will 
be compiled by the {\it Planck Surveyor} satellite (Blain 1998). A direct test 
of the properties of dust in the central regions of high-redshift ULIRGs will 
require a space-borne 
interferometer such as {\it SPECS} (Mather et al.\ 1998a,b). 
\end{enumerate} 
 
\section*{Acknowledgements} I thank Jean-Paul Kneib, Malcolm Longair, 
Priya Natarajan and an anonymous referee for their 
helpful comments on the manuscript, and PPARC and MENRT
for support. This research has 
made use of the NASA/IPAC Extragalactic Database (NED) which is operated by 
the Jet Propulsion Laboratory, California Institute of Technology, under 
contract with the National Aeronautics and Space Administration.

\end{document}